\documentclass[fleqn,10pt]{wlscirep}
\usepackage[utf8]{inputenc}
\usepackage[T1]{fontenc}
\usepackage{bm}
\usepackage{xr-hyper}
\usepackage{hyperref}
\AtBeginDocument{\let\hbar\hslash}

\title{Soft-Phonon-Driven Effective Inversion-Symmetry Crossover in Quantum Paraelectrics}

\author[1,$\dagger$,*]{Xiaojiang Li}
\author[2,$\dagger$]{Guodong Zhao}
\author[2,$\dagger$]{Fei Yang}
\author[4,5]{Seng Huat Lee}
\author[3]{Richard D. Schaller}
\author[1]{Jong-Woo Kim}
\author[1]{Matthew Krogstad}
\author[2]{Long-Qing Chen}
\author[1,*]{Philip J. Ryan}

\affil[1]{Advanced Photon Source, Argonne National Laboratory, Lemont, IL, USA}

\affil[2]{Department of Materials Science and Engineering and Materials Research Institute, The Pennsylvania State University, University Park, PA 16802, USA}

\affil[3]{Center for Nanoscale Materials, Argonne National Laboratory, Lemont, IL, USA}

\affil[4]{Department of Physics, The Pennsylvania State University, University Park, Pennsylvania 16802, USA}

\affil[5]{2D Crystal Consortium, Materials Research Institute, The Pennsylvania State University, University Park, Pennsylvania 16802, USA}

\affil[*]{Corresponding authors. E-mails: xiaojiang.li@anl.gov, pryan@anl.gov}

\begin{abstract}
Symmetry lays the foundation of condensed matter physics and its experimental manifestation provides fundamental insight into the collective behaviors of quantum materials. Optical second-harmonic generation (SHG) is widely regarded as a fingerprint of inversion-symmetry breaking, yet whether and how collective lattice dynamics govern the nonlinear optical manifestation of local inversion-symmetry breaking remains unknown. Here, combining optical SHG, diffuse X-ray scattering, and microscopic theory, we reveal a phonon-regulated mechanism governing the temperature-dependent manifestation of local inversion-symmetry breaking in quantum paraelectric material KTaO$_3$. We demonstrate that an oxygen-defect-mediated nonlinear optical channel is strongly coupled to the host soft polar mode, whose thermal fluctuations scramble the associated electronic phase coherence and thereby suppress the nonlinear manifestation of local inversion-symmetry breaking at elevated temperatures. Consequently, the nonlinear response exhibits a temperature-driven crossover from a regime in which local inversion-symmetry breaking is optically manifest to one that appears effectively centrosymmetric, without any accompanying structural change. Our findings revise the conventional picture of the temperature-dependent manifestation of inversion-symmetry breaking in quantum paraelectrics and establish a framework for understanding and engineering defect-mediated nonlinear optical responses in materials hosting low-energy polar excitations.

\end{abstract}
\begin{document}

\flushbottom
\maketitle

\thispagestyle{empty}


\section*{Introduction}

Defects are an inevitable component of real materials and often break crystal symmetries on a local scale without altering the global structure. Such local symmetry breaking can generate physical responses that are forbidden in the ideal crystal, providing a unique window into hidden microscopic degrees of freedom. However, whether the manifestation of these locally symmetry-broken regions is determined solely by their static structure, or can be regulated by collective excitations of the host lattice, remains largely unexplored.

Quantum paraelectric materials, such as KTaO$_3$ (KTO) and SrTiO$_3$ (STO), provide an ideal platform for addressing these questions. These materials are characterized by strong soft-phonon fluctuations associated with an incipient lattice instability. While conventional first-principles calculations suggest a low-temperature ferroelectric transition, zero-point lattice fluctuations prevent condensation of the soft mode and stabilize a quantum paraelectric ground state respecting inversion symmetry~\cite{rowley2014ferroelectric,Yang2026}. In general, optical second-harmonic generation (SHG) is symmetry-forbidden in centrosymmetric crystals. However, despite their globally centrosymmetric structures, pronounced SHG signals with acute temperature dependence are consistently observed in both KTO and STO, under both equilibrium conditions~\cite{derhorst1996variation,ojha2024quantum,vogt1990hyper,voigt1994experimental, uesu2004polar} and terahertz excitation~\cite{li2019terahertz,cheng2023terahertz,li2023terahertz}. The coexistence of a globally centrosymmetric crystal structure with a large second-order nonlinear optical response raises fundamental questions regarding the microscopic origin and temperature dependence of inversion-symmetry breaking.

The prevailing explanation attributes this pronounced temperature-dependent SHG response to polar nano-regions (PNRs), whose growth~\cite{Uwe_PhysRevB.33.6436,uwe1989ferroelectricSTO,ojha2024quantum,prusseit1990second,uesu2004polar}  and cooperative correlations upon cooling are proposed to generate macroscopic ferroelectric state and break global inversion-symmetry~\cite{aktas2014polar}. This framework has been widely invoked to explain the nonlinear response in both equilibrium~\cite{ojha2024quantum,derhorst1996variation,uesu2004polar,aktas2014polar} and non-equilibrium~\cite{li2019terahertz,cheng2023terahertz,li2023terahertz} conditions. However, several observations challenge this picture, including the weak temperature evolution of polar nano-region size inferred from Raman measurements (Supplement Material Sec.~\ref{sec_PNR})~\cite{Uwe_PhysRevB.33.6436,vogt1991evidence}, the increasing fragmentation of polar nano-regions observed in atomic-resolution imaging~\cite{zhang2025nanoscale}, and the report to reproduce key nonequilibrium SHG observations without invoking cooperative polar correlations or field-induced ferroelectricity~\cite{74d5-4hsw}. These observations motivate a re-examination of the microscopic mechanism governing the temperature-dependent nonlinear optical manifestation of local inversion-symmetry breaking in centrosymmetric quantum paraelectrics.

\begin{figure}[htb]
\begin{center}
\includegraphics[width=8.6cm]{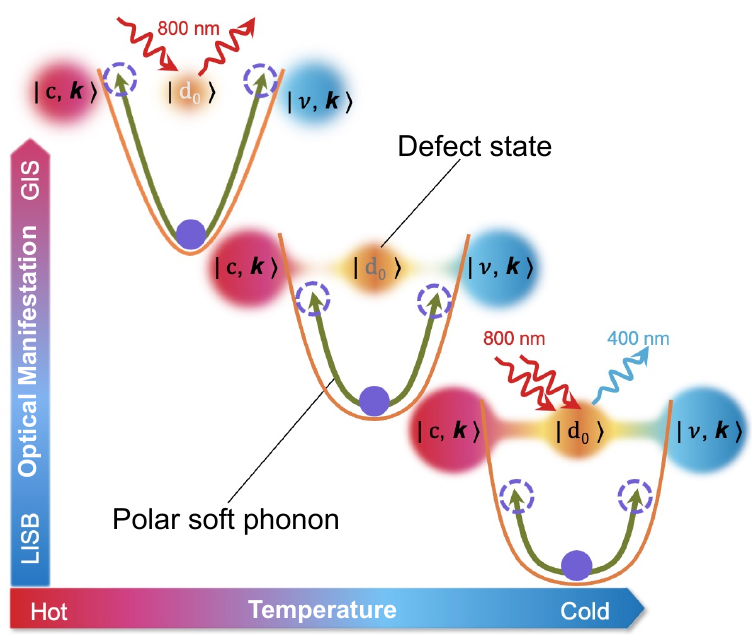}
\caption{Schematic illustration of the soft-phonon-driven effective inversion-symmetry crossover. In a globally centrosymmetric quantum paraelectric material, oxygen defects locally break inversion symmetry and introduce mid-gap states that mediate defect-assisted virtual electronic transitions in the second-order nonlinear optical response. At elevated temperatures, thermal fluctuations of the soft polar mode scramble the phase coherence of these defect-assisted transitions, strongly suppressing the nonlinear response. Upon cooling, the reduced thermal population of the soft mode progressively restores coherence and reveals the underlying defect-mediated nonlinear channel. The nonlinear response therefore crosses over continuously from a regime in which local inversion-symmetry breaking is present but effectively hidden to one in which it becomes macroscopically manifest, without a corresponding change in global crystallographic symmetry. GIS, global inversion symmetry; LISB, local inversion-symmetry breaking.}
\label{schematic}
\end{center}
\end{figure}

Combining optical SHG, diffuse X-ray scattering, and density functional theory (DFT) on structurally simpler KTaO$_3$ single crystal, we reveal a microscopic origin of the strong temperature-dependent nonlinear optical response in quantum paraelectrics. Optical SHG measurements indicate that the temperature-dependent manifestation of local inversion-symmetry breaking is primarily independent of the defect concentration, but intimately linked to thermal excitation of the soft polar mode. Diffuse X-ray scattering measurements and the corresponding  three-dimensional differential pair distribution function (3D-$\Delta$PDF) analysis further demonstrate that two-body correlations remain dominated by fluctuations of transverse phonons throughout the investigated temperature range. DFT calculations identify localized inversion-symmetry-breaking oxygen defects embedded within an otherwise centrosymmetric lattice as the microscopic source of the nonlinear response. These defects host localized electronic states that provide an intermediate channel for coherent virtual electronic transitions underlying the SHG response, while the host soft polar mode governs the coherence through strong dynamical coupling. As temperature increases, thermal fluctuations of the soft mode progressively scramble the electronic phase coherence associated with the nonlinear channel, thereby suppressing the optical manifestation of the underlying local symmetry. The resulting nonlinear response therefore exhibits a crossover from a regime in which local inversion-symmetry breaking is optically manifested to one in which it becomes optically hidden, despite the absence of any structural transition. The physical picture is summarized schematically in Fig.~\ref{schematic}. The pronounced temperature dependence of SHG therefore reflects a previously unrecognized mechanism by which collective lattice fluctuations regulate the macroscopic manifestation of local inversion-symmetry breaking.

\section*{Results}

\subsection*{Temperature-dependent manifestation of local inversion-symmetry breaking}
\label{optical shg measurements}
\begin{figure}[htb]
\begin{center}
\includegraphics[width=15.6cm]{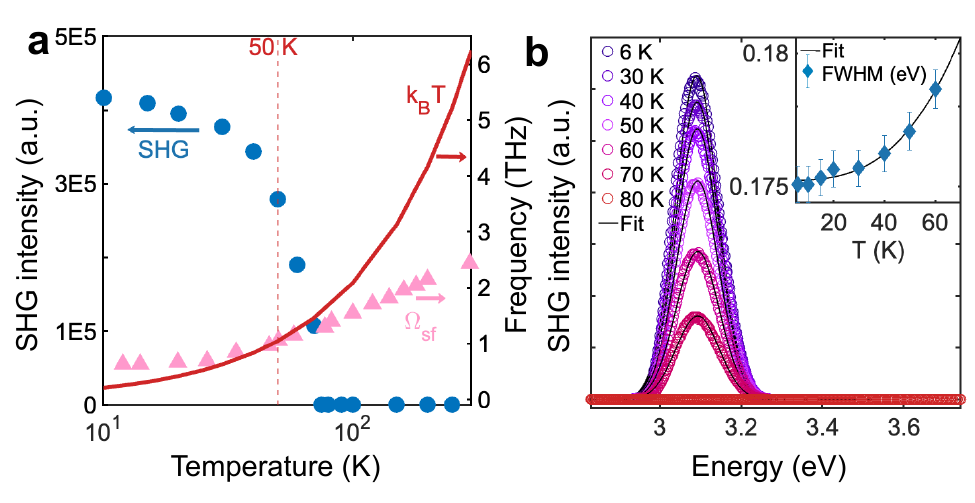}
\caption{
Temperature-dependent SHG measurements on KTaO$_3$.
(a) SHG intensity (blue solid circles), thermal excitation energy $k_B T$ (red line), and soft transverse optical phonon energy (pink solid triangles, extracted from Ref.~\citenum{yukiichikawa:092106}) as a function of temperature. The red dashed line marks $T = 50~\mathrm{K}$, where the thermal excitation energy becomes comparable to the soft-mode energy and the SHG intensity starts to saturate.
(b) Temperature-dependent SHG spectral lineshape together with Gaussian fits. Inset: extracted full width at half maximum (FWHM) as a function of temperature, fitted by $f(T)=aT^3+b$. Error bars on the extracted FWHM values represent $95\% $ confidence intervals obtained from the Gaussian fits.}
\label{fig1}
\end{center}
\end{figure}

The temperature-dependent optical second-harmonic generation response of KTaO$_3$ is summarized in Fig.~\ref{fig1}. A pronounced SHG signal emerges below $75~\mathrm{K}$, increases abruptly by more than three orders of magnitude upon cooling and saturates around $50~\mathrm{K}$ [Fig.\ref{sup_PNR}]. Above $75~\mathrm{K}$, the SHG signal falls below the detection threshold under identical excitation conditions, but can be recovered under substantially stronger optical pumping. For example, at $110~\mathrm{K}$, a laser fluence nearly three orders of magnitude larger is needed to yield an SHG response comparable to that observed at $75~\mathrm{K}$ [Fig.~\ref{sup_SHG_powerdependence}]. Despite this pronounced enhancement of SHG, no evidence for a corresponding structural phase transition is detected, as will be established in the diffuse X-ray scattering measurement presented in the following section. Such a dramatic increase of SHG signal in the absence of any detectable structural phase transition is unexpected for a macroscopically cubic and globally centrosymmetric crystal, where SHG is forbidden within the electric-dipole approximation.

Additional insight is provided by the SHG spectral lineshape [Fig.~\ref{fig1}(b)]. The spectra are well described by a pure Gaussian profile; incorporating  a Lorentzian contribution via a Voigt fitting yields negligible improvement, suggesting that the temperature-dependent linewidth is dominated by thermally induced inhomogeneous broadening on top of excitation laser bandwidth. Notably, the full width at half maximum (FWHM) of SHG scales approximately as $T^3$, reminiscent of the acoustic-phonon-induced dephasing commonly observed in localized solid-state emitters~\cite{xue2020single,sontheimer2017photodynamics,lienhard2016bright} and anharmonic phonon broadening revealed by Raman spectroscopy\cite{menendez1984temperature}. In contrast, the saturation temperature of the SHG response closely coincides with the temperature scale associated with thermal excitation of the soft polar mode [Fig.~\ref{fig1}(a)]. This dichotomy suggests that the linewidth broadening and the temperature dependence of SHG intensity may reflect distinct roles of lattice dynamics: the $T^3$ linewidth broadening is consistent with acoustic-phonon-induced dephasing, whereas the saturation temperature of the nonlinear response points to a possible connection with the excitation of the low-energy soft polar mode.

This picture is further constrained by the dependence of the SHG intensity on the concentration of oxygen vacancies, an omnipresent defect class in transition-metal oxides~\cite{derhorst1996variation,yamaichi1987photoluminescence,kapphan2005uv,betzler1980second,grabner1969photoluminescenceSTO}. As shown in Fig.~\ref{sup_sample_anneal},  high-vacuum annealing of a pristine sample yields a robust enhancement of the SHG intensity, whereas oxygen-atmosphere annealing drastically suppresses it, tracking the expected evolution of oxygen vacancy density~\cite{ojha2024quantum,derhorst1996variation,voigt1994experimental,vogt1990hyper}. These trends confirm that defect concentration directly gates the SHG intensity. Crucially, however, the characteristic onset temperature of SHG remains invariant under disparate annealing treatments. This further decoupling indicates that while the concentration of static defects dictates the absolute magnitude of the nonlinear response at low temperature, the underlying coherent nonlinear process is controlled by a distinct lattice degree of freedom that is largely insensitive to oxygen-vacancy concentration.

Taken together, the SHG measurements establish four key observations: (i) the nonlinear response develops a pronounced enhancement below a well-defined temperature scale without any detectable structural transition, (ii) the saturation temperature of SHG response coincides with the temperature scale associated with thermal excitation of the soft polar mode , (iii) the response amplitude is strongly controlled by oxygen-vacancy concentration, and (iv) the characteristic onset temperature scale remains insensitive to defect density. These findings indicate that the microscopic source of inversion-symmetry breaking and the mechanism governing its temperature-dependent manifestation are distinct. Identifying the lattice degrees of freedom responsible for this behavior therefore requires direct characterization of the underlying structural correlations, which we address through diffuse X-ray scattering measurements.



\subsection*{Phonon-driven displacement correlations}

\begin{figure}[htb]
\begin{center}
\includegraphics[width=15.7cm]{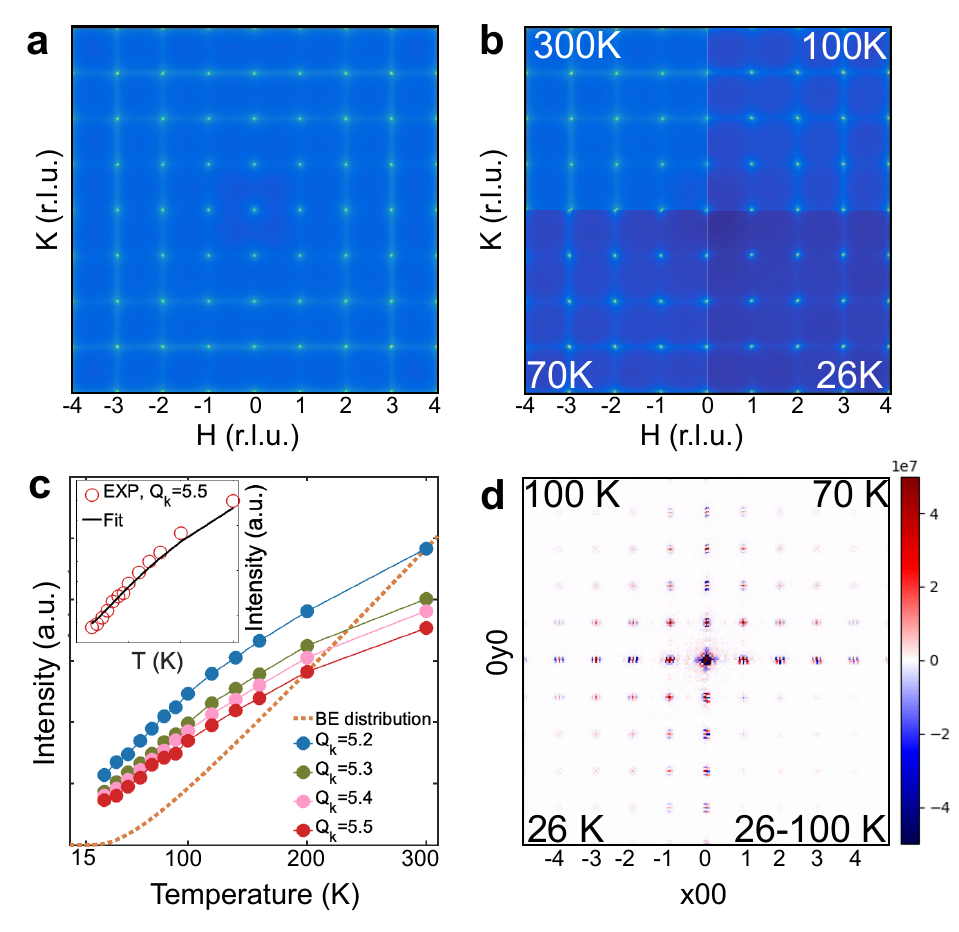}
\caption{
Temperature-dependent diffuse X-ray scattering measurements in KTaO$_3$.
(a) Symmetrized diffuse X-ray scattering intensity in the $(001)$ plane at $300~\mathrm{K}$. The scattering is dominated by diffuse planes perpendicular to the principal reciprocal-lattice axes, with extinction of planes intersecting the direct-beam center. (b) Diffuse scattering in the $(001)$ plane at  $300~\mathrm{K}$, $100~\mathrm{K}$, $70~\mathrm{K}$, and $26~\mathrm{K}$. No superlattice peaks, Bragg-peak splitting or other anomalous intensity changes are observed upon cooling. (c) Temperature dependence of the diffuse scattering intensity at selected momentum transfers $\mathbf{Q}_k$ along the $[100]$ direction. The orange dotted curve shows the Bose–Einstein occupation expected for a single harmonic mode with fixed mode frequency, which does not reproduce the measured temperature dependence. Inset: measured diffuse intensity at $Q_k = 5.5$~r.l.u. together with the fit of phonon-population model described in the main text. (d) Temperature evolution of the 3D-$\Delta$PDF reconstructed from the diffuse scattering data at $100~\mathrm{K}$, $70~\mathrm{K}$, and $26~\mathrm{K}$. The fourth quadrant of the figure shows the differences between~$26~\mathrm{K}$ and $100~\mathrm{K}$. 
}
\label{fig2}
\end{center}
\end{figure}

To understand the lattice correlations underlying the anomalous SHG response, we conducted diffuse X-ray scattering. This technique provides a sensitive probe of two-body correlations, making it a powerful tool for investigating orders of defects and structures in complex materials~\cite{osborn2025diffuse,krogstad2020reciprocal,he2025resolving,kopecky2012x}. In KTaO$_3$, the diffuse scattering pattern is dominated by a set of planes perpendicular to the principal reciprocal-lattice axes at integer coordinates [Fig.~\ref{fig2}(a,b)]. The diffuse intensity decreases monotonically upon cooling [Fig.~\ref{fig2}(b,c)], but scales positively with momentum transfer $\mathbf{Q}$ [Fig.~\ref{sup_diffuse_xray_scattering}], which are the characteristic behaviors of thermal diffuse scattering~\cite{xu2005determination,sangiorgio2018correlated}. Importantly, no additional signatures, including superlattice peaks, Bragg peak splitting or anomalous intensity changes are observed throughout the entire temperature range of experiment [Fig.~\ref{fig2}(b)]. These observations indicate that the dramatic enhancement of SHG occurs without the emergence of a detectable symmetry-lowering structural transition.

A striking feature of the diffuse scattering pattern is that the full diffuse plane extincts when it intersects the direct-beam center, namely, planes satisfying $H$, $K$, or $L=0$ [Fig.~\ref{fig2}(a)]. Such a selection rule places strong constraints on the underlying displacement correlations of thermal diffuse scattering. Expanding the total scattering intensity in powers of the relative atomic displacement ${\bf e}_{jj'}={\bf u}_j-{\bf u}_{j'}$~\cite{welberry1994interpretation},
\begin{equation}
I \propto \sum_{jj'} b_j b_{j'}
e^{i\mathbf{Q}\cdot(\mathbf{R}_j-\mathbf{R}_{j'})}
\big[
1+i\mathbf{Q}\cdot{\bf e}_{jj'}
-\frac{1}{2}
(\mathbf{Q}\cdot{\bf e}_{jj'})^2+\cdots
\big],
\end{equation}
where $b_j$ denotes the scattering length and $\mathbf{Q}=\mathbf{G}+\mathbf{q}$ is the scattering vector, shows that the leading diffuse contribution, which is the first-order thermal diffuse scattering arising from two-body correlations, disappears when $\mathbf{Q}\cdot{\bf e}_{jj'}=0$. The observed extinction of the full diffuse plane therefore demonstrates that the atomic displacements are strictly perpendicular to the whole scattering plane, consistent with transverse-phonon-like fluctuations observed from neutron diffuse scattering\cite{he2025resolving}.

To further elucidate the correlations revealed by diffuse X-ray scattering, we examined the temperature dependence of the diffuse intensity at various $\mathbf{Q}$ positions [Fig.~\ref{fig2}(c)]. The diffuse intensity decreases smoothly and diminishes more rapidly upon cooling , contrary to the expectation of progressively enhanced static polar correlations. Instead, this behavior is naturally captured by a phonon-population model incorporating mode softening, a hallmark of quantum paraelectric behavior. In KTaO$_3$, only two low-energy transverse  branches remain thermally populated below $270~\mathrm{K}$~\cite{perry_PhysRevB.39.8666,axe1970anomalous}, and the first-order thermal diffuse scattering intensity from phonons is proportional to the Bosonic  occupation~\cite{xu2005determination}. However, the reduction in phonon population with cooling alone does not fully account for the observed temperature dependence [Fig. \ref{fig2}(c)].  To address this, we focused on the diffuse intensity at $X$ point along  $\Delta$ direction, where phonon frequencies are well characterized by neutron scattering~\cite{perry_PhysRevB.39.8666,axe1970anomalous}. By incorporating mode softening, the phonon-population model successfully captures the main features of the measured temperature-dependent diffuse intensity [inset of Fig.~\ref{fig2}(c)].

In order to obtain more intuitive understanding of the diffuse X-ray scattering data in real space, we calculated the three-dimensional differential pair distribution function (3D-$\Delta$PDF), which provides a direct measure of the probability distribution of interatomic vectors that deviate from those of the average structure~\cite{weber2012three}. The resulting maps across the investigated temperature range exhibit positive peaks at average interatomic distances and negative peaks at nearest neighbors [Fig.~\ref{fig2}(d) and Fig. \ref{sup_detaPDF}]. These features are characteristic signatures of acoustic-phonon correlations with in-phase atomic displacements~\cite{sangiorgio2018correlated}. 

Combining the diffuse X-ray scattering and 3D-$\Delta$PDF analyses, we find that the lattice correlations of KTO throughout the investigated temperature range remain dominated by phonon fluctuations, with the strongest contribution arising from TA modes and a significant temperature-dependent contribution from the soft TO$_1$ branch. Importantly, neither the reciprocal-space diffuse scattering nor the real-space 3D-$\Delta$PDF maps exhibit signatures of an emerging ordered phase or a qualitative enhancement in the underlying displacement correlations upon cooling. The pronounced enhancement of SHG therefore occurs within a lattice environment that remains dynamically fluctuating without developing a new static symmetry-broken phase. These observations establish collective phonon fluctuations as the dominant source of lattice correlations in KTO and provide the experimental foundation for understanding how lattice dynamics regulate the temperature-dependent manifestation of local inversion-symmetry breaking.

\subsection*{Microscopic mechanism of phonon-regulated nonlinear response}

\begin{figure}[htb]
\begin{center}
\includegraphics[width=15.5cm]{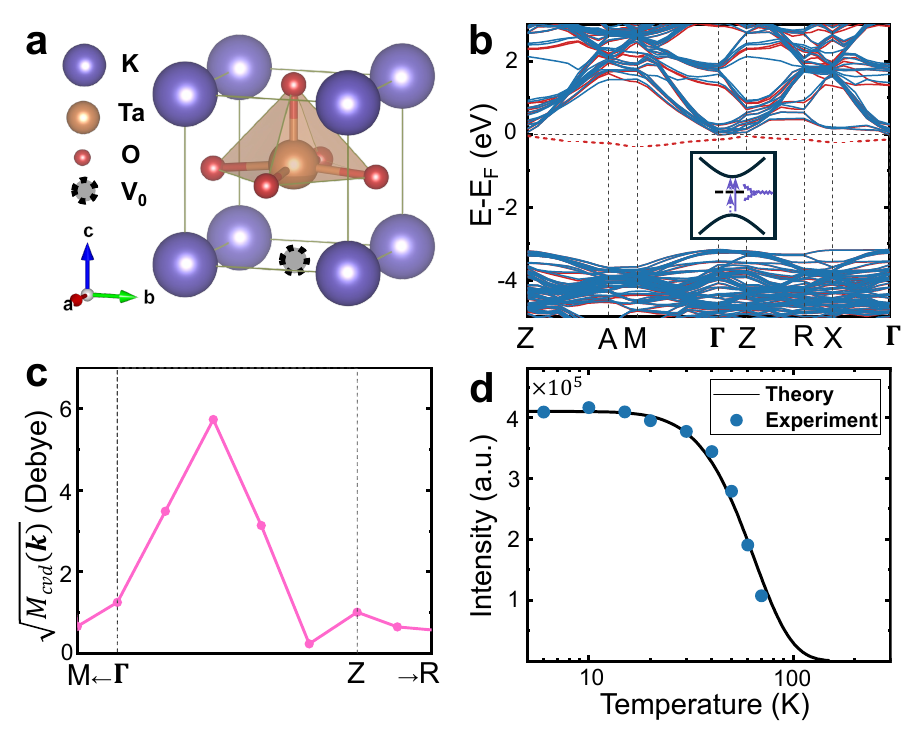}
\caption{
Defect-assisted mechanism for the temperature-dependent SHG response in KTaO$_3$.
(a) Atomic structure of KTO containing a single charged oxygen vacancy $V_{\mathrm O}^{+}$ which breaks local inversion symmetry.
(b) Hybrid-functional band structure of a $2\times2\times4$ KTO supercell containing a single oxygen vacancy. Red and blue solid curves denote the two spin channels, while the dashed curve indicates the in-gap defect state. Inset: schematic illustration of the defect-assisted two-step optical transition between valence and conduction bands under 800~nm excitation.
(c) Momentum-resolved defect-assisted nonlinear transition strength near the $\Gamma$-$Z$ direction, obtained from the combined valence-to-defect and defect-to-conduction dipole matrix elements.
(d) Calculated temperature dependence of the SHG intensity compared with experiment.
}
\label{fig3}
\end{center}
\end{figure}

Motivated by previous hybrid-functional studies identifying charged oxygen vacancies as intrinsic carrier-trapping centers in KTaO$_3$~\cite{modak2021energetic}, we consider the singly charged oxygen vacancy $V_{\rm O}^{+}$ as a representative model for optically active defect configurations. As shown in Fig.~\ref{fig3}(a,b), hybrid functional (HSE06) calculations~\cite{HSE06-1,HSE06-2,HSE06-3,Kresse1996-1,Kresse1996-2} reveal a localized in-gap defect state $|d_0\rangle$, which breaks the local inversion symmetry and provides an electronic pathway for second-order nonlinear optical process that is forbidden in the ideal crystal. 

Under 800~nm excitation, this localized state serves as an intermediate channel that mediates coherent \emph{virtual} electronic transitions between the valence-band states $|v,\mathbf{k}\rangle$ and  the conduction-band states $|c,\mathbf{k}\rangle$. The parametric process can be represented schematically as:
\begin{equation}
|v,\mathbf{k}\rangle
\rightarrow
|d_0\rangle
\rightarrow
|c,\mathbf{k}\rangle
\rightarrow
|v,\mathbf{k}\rangle.
\end{equation}

Integrating out the local defect state yields an effective second-order nonlinear optical coupling:
\begin{equation}
    \left(H_{\rm eff}\right)_{vc}^{(2)}
  =M_{cvd}(\mathbf{k})\Lambda({\bf k})E^2.
\end{equation}
Here $
M_{cvd}(\mathbf{k})$ is the defect-assisted optical matrix element; $\Lambda({\bf k})$is the virtual-state energy denominator obtained from the Schrieffer-Wolff transformation~\cite{PhysRev.149.491}. This defect-assisted second-order optical pathway originates from coherent virtual electronic transitions via the localized defect state rather than a static classical polarization field. The momentum dependence of $M_{cvd}(\mathbf{k})$ evaluated from the HSE06 calculations is shown in Fig.~\ref{fig3}(c), establishing a finite, non-vanishing nonlinear optical vertex over an extended region of momentum space.

The temperature dependence arises because the localized defect state is embedded in a fluctuating polar phonon environment, consistent with the diffuse X-ray scattering measurements that the lattice correlations in KTaO$_3$ remain dominated by phonon fluctuations throughout the investigated temperature range. Instead of acting as rigid static defects, the oxygen vacancies remain weakly pinned and highly susceptible to local polar fluctuations~\cite{Ouhbi2021,Ojha2021divac}. Consequently, in the presence of local ionic displacements associated with soft-phonon normal coordinates $\{Q_\nu\}$, the defect-assisted optical matrix element is sensitive to the environment and becomes configuration dependent,
\begin{equation}
M_{cvd}(\mathbf{k};\{Q_\nu\})
=
\langle c,\mathbf{k}|
\hat d
|d[\{Q_\nu\}]\rangle
\langle d[\{Q_\nu\}]|
\hat d
|v,\mathbf{k}\rangle .
\end{equation}
For weak lattice fluctuations, the primary effect of the soft polar mode is to introduce fluctuating phases into the defect-mediated nonlinear polarization channel via local distortions, since long-wavelength lattice displacements constitute perturbations that are locally equivalent to translations of the underlying lattice environment, whose leading-order effect is a phase modulation of the nonlinear polarization response. This allows us to parameterize the matrix element as $M_{cvd}(\mathbf{k};\{Q_\nu\})
=
M_{cvd}(\mathbf{k})
e^{-i\phi(\{Q_\nu\})}$, where the phase is given by $
\phi(\{Q_\nu\})
=
\sum_\nu g_\nu Q_\nu$~\cite{kittel1963quantum}, with $g_\nu$ denoting the effective defect-phonon coupling constant. The nonlinear response is determined by the coherent thermal average of this matrix element, and the fluctuations of the normal coordinates $Q_\nu$ is Gaussian based on the assumption of harmonic phonons. By using a second-order cumulant expansion, the coherent expectation value can be written as:
\begin{equation}
\overline{M}_{cvd}(\mathbf{k},T)
=
M_{cvd}(\mathbf{k}) e^{-W(T)},
\label{eq:Mrenorm}
\end{equation}
where the Debye--Waller-like phase-decoherence factor is determined by $
W(T)
=
\frac{1}{2}
\sum_\nu
|g_\nu|^2
\left\langle
Q_\nu^2
\right\rangle_T$. For a harmonic mode of frequency $\Omega_\nu$~\cite{abrikosov2012methods,mahan2013many}, $
\left\langle
Q_\nu^2
\right\rangle_T
=
\frac{\hbar}{2\Omega_\nu}
\left[
2n_B(\Omega_\nu,T)+1
\right]$, where $n_B(\Omega_\nu,T)$ is the Bose--Einstein distribution function. Because the transverse optical soft mode in KTO has an anomalously small frequency~\cite{74d5-4hsw,li2023terahertz,cheng2023terahertz}, its displacement fluctuations are strongly amplified and dominate the phase-decoherence factor $W(T)$. 
As a result, thermal soft-phonon fluctuations strongly suppress the coherent nonlinear optical vertex at elevated temperatures, whereas cooling progressively restores it. Detailed derivations of the effective model and a comprehensive analysis of the parity constraints governing the dipole transition matrix elements are presented in Supplementary Section~\ref{sec:calculations}.

Within the framework of the semiconductor optical Bloch equations~\cite{Haug1994}, the coherent second-harmonic polarization can then be written as:
\begin{equation}
P^{(2)}_e(2\omega;T)
    =
    E_\omega^2
    \sum_{\mathbf{k}}
    \frac{
    d_{vc}(\mathbf{k})
    M_{cvd}(\mathbf{k})
    e^{-W(T)}
    \Lambda(\mathbf{k})
    }{
    \varepsilon_c(\mathbf{k})
    -
    \varepsilon_v(\mathbf{k})
    -
    2\hbar\omega
    -
    i\hbar\gamma_{\mathbf{k}}(T)
    }.
\label{eq:model}
\end{equation}
with the measured SHG intensity scales as $
I_{2\omega}(T)
\propto
|P_e(2\omega;T)|^2$. The dephasing rate  $\gamma_k(T)$ in the optical denominator is modeled phenomenologically~\cite{Haug1994,PhysRevB.93.235433,PhysRevB.92.155414,Wu2010SpinDynamics}, incorporating the experimentally observed $T^3$ broadening of the SHG linewidth, consistent with acoustic-phonon-induced dephasing discussed above. In contrast, the temperature dependence of the SHG intensity is governed primarily by the fluctuation-induced factor $e^{-W(T)}$, which is dominated by the soft polar mode. Using the fluctuation-induced coherence factor together with experimentally determined linewidth , the model successfully reproduces the measured temperature dependence of the SHG intensity [Fig.~\ref{fig3}(d)].

The resulting picture naturally explains the factorization observed experimentally. Oxygen vacancies provide the local inversion-symmetry breaking and establish the defect-mediated nonlinear optical channel, thereby determining the overall magnitude of the SHG response. Consistent with the diffuse-scattering and SHG measurements, the fluctuations of soft polar mode modulate temperature-dependent lattice correlations and  regulates the coherence and manifestation of this channel without altering the underlying global symmetry. The pronounced enhancement of SHG upon cooling therefore reflects the progressive quenching of fluctuation-induced decoherence rather than the emergence of a new symmetry-broken phase.

\section*{Discussion}

Our work reveals that symmetry and its experimental manifestation need not be in one-to-one correspondence. In KTaO$_3$, local inversion-symmetry-breaking defects persist throughout the investigated temperature range, while the crystal remains globally centrosymmetric. Yet the nonlinear optical response evolves from an effectively centrosymmetric regime at elevated temperatures to a defect-dominated inversion-symmetry-breaking regime upon cooling. This evolution is not driven by a change in crystallographic symmetry, but by the temperature-dependent thermal fluctuations of the soft polar mode. The fluctuations of this mode scramble the phase coherence of the virtual electronic transitions responsible for SHG at elevated temperatures, whereas cooling progressively restores coherence and reveals the underlying defect-mediated nonlinear channel. The apparent emergence of inversion-symmetry breaking therefore reflects a dynamical crossover in the nonlinear response rather than a change in the underlying symmetry. Because this mechanism requires only localized inversion-symmetry-breaking defects and low-energy polar lattice excitations, ingredients found across a broad class of  defect-bearing polarizable materials, our results suggest that fluctuation-regulated manifestations of local symmetry breaking may be more widespread than previously recognized.

More broadly, our results establish that collective excitations can regulate not the existence of broken local symmetry itself, but its experimental manifestation. This perspective shifts the focus from static symmetry breaking alone to the dynamical processes that govern how symmetry is manifested in measurable observables. The framework therefore provides a general route for understanding defect-mediated nonlinear optical phenomena and suggests new opportunities to control symmetry-sensitive optical responses through collective-mode engineering, including temperature tuning, strain control, and coherent phonon excitation, without requiring structural symmetry breaking.




\vskip 0.2cm
\noindent{\bf Acknowledgments}\\
This research utilized the resources of the Advanced Photon Source, a US Department of Energy (DOE) Office of Science user facility at Argonne National Laboratory, and was supported by the U.S. DOE Office of Science under Contract No. DE-AC02-06CH11357. This work was supported by Laboratory Directed Research and Development (LDRD) funding from Argonne National Laboratory, provided under Contract No. DE-AC02-06CH11357 by the Director, Office of Science, US DOE. R.D.S. acknowledges support for optical spectroscopy work performed at the Center for Nanoscale Materials, a U.S. Department of Energy Office of Science User Facility, under Contract No. DE-AC02-06CH11357. G.D.Z., F.Y., and L.Q.C. acknowledge support from the U.S. Department of Energy, Office of Science, Basic Energy Sciences, under Award No. DE-SC0020145 as part of the Computational Materials Sciences Program. F.Y., and L.Q.C. also acknowledge support from the Donald W. Hamer Foundation through a Hamer Professorship at Penn State.

\vskip 0.2cm
\noindent{\bf Author Contributions}\\
X.J.L. and P.J.R. conceived the study. X.J.L. and R.D.S. performed the SHG measurements. M.K.  and X.J.L. performed the diffuse X-ray scattering measurements. G.D.Z. and F.Y. carried out the theoretical calculations. X.J.L., F.Y. and G.D.Z wrote the original draft of the manuscript. All authors contributed to the interpretation of the results and revision of the manuscript. 

\bibliography{sample}

\end{document}


\flushbottom
\maketitle
\tableofcontents

\thispagestyle{empty}

\section{Sample}
\label{Sample}
\begin{figure}[htb]
\begin{center}
\centering
  {\includegraphics[width=8cm]{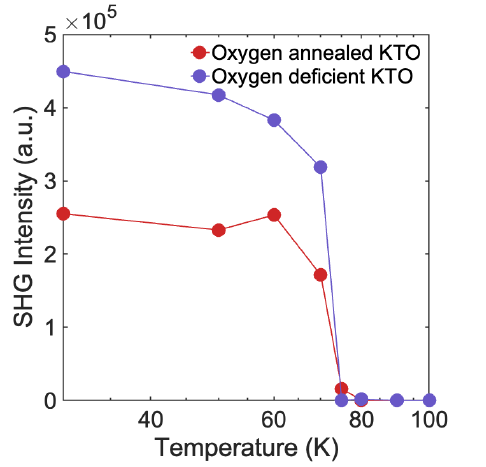}}
  \caption{Effect of oxygen annealing on the temperature-dependent SHG response of KTaO$_3$. SHG intensity as a function of temperature for the oxygen-deficient sample before and after annealing in an oxygen atmosphere at $900~^\circ$C for 3~hours. Oxygen annealing suppresses the low-temperature SHG intensity by more than $40\%$, while the characteristic temperature scale of the response remains essentially unchanged.}
  \label{sup_sample_anneal}
\end{center}
\end{figure}

The KTaO$_3$ samples used in this work are double-side polished single crystals with (001) surfaces purchased from MTI. Pristine samples exhibited only weak, near-background SHG responses under the measurement conditions used here. The oxygen-deficient sample was prepared following the procedure described in Ref.~\citenum{ojha2024quantum,Ojha2021divac}, and its temperature-dependent response is presented in the main text and Fig.~\ref{sup_sample_anneal}. To modify the oxygen-vacancy concentration, the oxygen-deficient sample was subsequently annealed in an oxygen atmosphere at $900~^\circ$C for 3 hours. Following annealing, the low-temperature SHG intensity was reduced by more than $40 \%$, consistent with a substantial contribution from oxygen-vacancy-related defect states. In contrast, the characteristic temperature scale of SHG response remains essentially unchanged [Fig.~\ref{sup_sample_anneal}]. These observations indicate that oxygen-vacancy concentration strongly influences the magnitude of the nonlinear response, while the characteristic temperature scale is governed by a distinct degree of freedom that is comparatively insensitive to oxygen-vacancy concentration. As discussed in the main text, the combined experimental and theoretical results identify the thermally populated soft polar mode as this intrinsic regulator.

\clearpage

\section{Optical second-harmonic generation measurements}
The optical second-harmonic generation measurements were performed using a Ti:sapphire pulsed laser with a central wavelength of $800~\mathrm{nm}$,  pulse duration $100~\mathrm{fs}$  and a repetition rate of $2~\mathrm{kHz}$ in a transmission geometry. The laser power incident on the sample was approximately $50~\mu\mathrm{W}$, and the laser spot diameter was estimated to be $96~\mu\mathrm{m}$ by using an optical pinhole method. After the sample, the directly transmitted beam was filtered by a $400~\mathrm{nm}$ band-pass filter to remove the $800~\mathrm{nm}$ beam. The sample was mounted in a Janis cryostat with temperature range from $5~\mathrm{K}$ to room temperature. The SHG spectrum was detected by using an HR300 spectrometer from Princeton Instruments.

\begin{figure}[htb]
\begin{center}
\centering
  {\includegraphics[width=8cm]{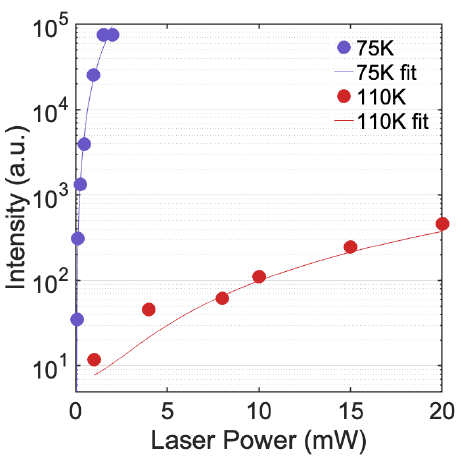}}
  \caption{
  Laser-power dependence of the SHG intensity measured at $75~\mathrm{K}$ and $110~\mathrm{K}$. The solid dots represent the experimental data, while the solid lines are quadratic fits. The SHG intensity is plotted on a logarithmic scale.
  }
  \label{sup_SHG_powerdependence}
\end{center}
\end{figure}

Figure~\ref{sup_SHG_powerdependence} shows the laser-power dependence of the SHG intensity measured at $75~\mathrm{K}$ and $110~\mathrm{K}$. At both temperatures, the SHG intensity exhibits a quadratic dependence on the incident laser power, consistent with a second-order nonlinear optical process. Under the standard excitation conditions used for the temperature-dependent measurements, with an incident laser power of approximately $50~\mu\mathrm{W}$, a pronounced SHG response is observed at $75~\mathrm{K}$. By contrast, at $110~\mathrm{K}$, excitation powers in the tens-of-milliwatts range are required to generate a comparable SHG intensity, corresponding to an increase of nearly three orders of magnitude in incident power. Power-dependent measurements over the corresponding excitation ranges were repeated at multiple temperatures, and no clear sample damage was observed.

\clearpage

\section{Comparison of SHG with polar nano-region size and dielectric permittivity}
\label{sec_PNR}
\begin{figure}[htb]
\begin{center}
\centering
  {\includegraphics[width=16cm]{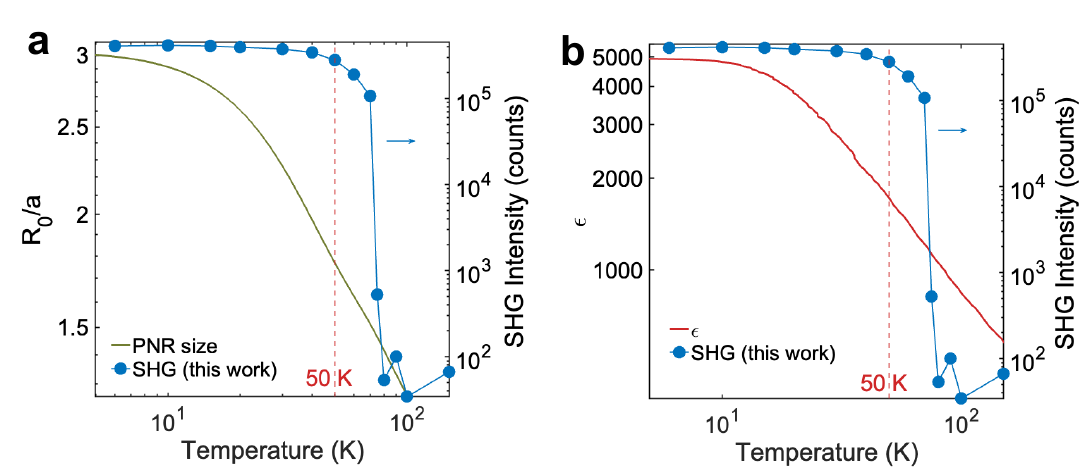}}
  \caption{Comparison of the temperature-dependent SHG response with PNR size and dielectric permittivity.
(a) Characteristic PNR size $\mathbf{R_0}/a$ and SHG intensity as functions of temperature. $\mathbf{R_0}$ denotes the PNR diameter and $a$ the lattice constant of KTaO$_3$. (b) Dielectric permittivity $\epsilon$ and SHG intensity as functions of temperature. PNR-size and permittivity data are extracted from Refs.~\citenum{Uwe_PhysRevB.33.6436} and ~\citenum{aktas2014polar}. The red dashed line marks $50~\mathrm{K}$, near the temperature at which the SHG response approaches saturation. }
  \label{sup_PNR}
\end{center}
\end{figure}

Upon cooling, the characteristic size of polar nano-regions (PNRs) of KTaO$_3$ expands from roughly 1.3 unit cells at $100~\mathrm{K}$ to about 3 unit cells at $2~\mathrm{K}$, as shown in Fig.~\ref{sup_PNR}(a). The temperature-dependent SHG response is plotted alongside for direct comparison. Crucially, from $90~\mathrm{K}$ to $50~\mathrm{K}$ -- the temperature at which the response begins to saturate -- the SHG intensity scales upward by more than three orders of magnitude. Over this identical temperature window, however, the nominal PNR size only increases from 1.33 to 1.76 unit cells. To estimate the enhanced SHG intensity from the growth of the PNR size by assuming the polarization is proportional to the volume of PNR, we have $I_{50K}\approx (1.76/1.33)^6*I_{90K} = 5.43*I_{90K}$. This calculated increase falls critically short of the experimentally observed three-order-of-magnitude surge, demonstrating that simple physical growth of the PNRs cannot alone explain the enhanced nonlinear optical response.

Another possible contribution to the temperature-dependent SHG response is the dielectric renormalization accompanying the softening of the polar mode. As shown in Fig.~\ref{sup_PNR}(b), however, the dielectric permittivity evolves continuously upon cooling, in marked contrast to the abrupt three-orders-of-magnitude increase of the SHG intensity within a narrow temperature interval. This comparison suggests that the observed SHG enhancement cannot be understood solely from dielectric renormalization either and instead points to an additional microscopic mechanism governing the nonlinear optical response.

\clearpage

\section{Diffuse X-ray scattering}
Three-dimensional volumes of diffuse X-ray scattering were collected at beamline 6-ID-D of the Advanced Photon Source (APS) using an incident X-ray energy of $90~\mathrm{keV}$ and a Dectris Pilatus 2M detector equipped with a 1-mm-thick CdTe sensor layer. Measurements were performed from $26~\mathrm{K}$ to $300~\mathrm{K}$ using an Oxford N-Helix Cryocooler, with flowing He gas below $100~\mathrm{K}$ and flowing N$_2$ gas above $100~\mathrm{K}$.

During data collection, the sample was continuously rotated about an axis perpendicular to the incident beam at a rate of $1^{\circ}\,\mathrm{s}^{-1}$ over a total angular range of $365^{\circ}$, while detector images were recorded every $0.1~\mathrm{s}$. For each sample and temperature, three sets of rotational scans were collected to fill the gaps between detector chips. The resulting diffraction images were stacked into a three-dimensional array, oriented using the NXRefine Python package, and transformed into reciprocal-space coordinates using the software package CCTW (Crystal Coordinate Transformation Workflow). This procedure enabled the determination of $S(\mathbf{Q})$ over a reciprocal-space range of approximately $\pm 15~\text{\AA}^{-1}$ in all directions.

\begin{figure}[htb]
\begin{center}
\centering
  {\includegraphics[width=8cm]{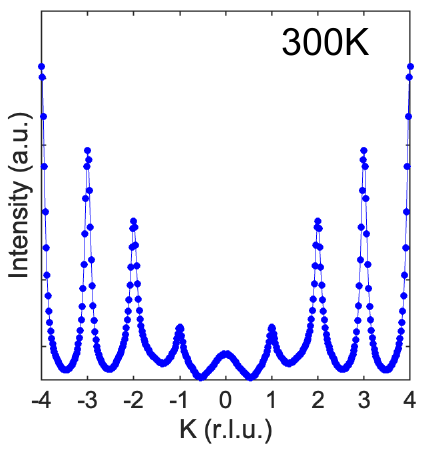}}
  \caption{ Momentum-transfer dependence of the diffuse X-ray scattering intensity at $300~\mathrm{K}$ . Intensity profile extracted from a reciprocal-space slice at $H = 2.5$ r.l.u. and integrated along the $L$  direction. The diffuse-scattering intensity generally increases with increasing $|\mathbf{Q}|$. }
  \label{sup_diffuse_xray_scattering}
\end{center}
\end{figure}

As shown in Fig.~\ref{sup_diffuse_xray_scattering}, the diffuse-scattering intensity generally increases with increasing momentum transfer $|\mathbf{Q}|$, consistent with the $\mathbf{Q}$-dependent scattering amplitude expected for phonon-mediated thermal diffuse scattering.

\clearpage

\section{Phonon-population model}
To better understand the temperature-dependent diffuse X-ray scattering, we introduced a simple phonon-population model based on the assumption that the temperature dependence of thermal diffuse scattering intensity solely determined by the thermal population of phonons. For the first-order thermal diffuse scattering, which usually dominates the diffuse scattering intensity at modest or low temperatures, it is linearly proportional to the phonon population~\cite{xu2005determination}, therefore the intensity of diffuse X-ray scattering is proportional to the population of phonons. Although the model is phenomenological and neglects many microscopic details, it captures the temperature dependence of diffuse scattering intensity remarkably well with only one global fitting parameter.

The model is written as
\begin{align}
    I(T)
    =
    C'
    \left[
    \frac{n(\omega_1,T)+1/2}{\omega_1}
    +
    \frac{n(\omega_2(T),T)+1/2}{\omega_2(T)}
    \right],
    \label{eq:toymodel}
\end{align}
where $C'$ is the global tuning parameter assumed to be temperature-independent. $n(\omega,T)$ is the Bose-Einstein distribution function, representing the thermal population of phonons.

For the temperature-dependent diffuse X-ray scattering near the $X$ point along the $\Delta$ direction in reciprocal space, neutron scattering measurements show that the TA phonon mode softens continuously upon cooling, while the TO$_1$ mode remains nearly temperature independent in this region. In the model, accordingly, we set the TO$_1$ phonon frequency near the $X$ point to be $5.39~\mathrm{THz}$ \cite{perry_PhysRevB.39.8666} and for the TA mode, we use the experimentally reported frequencies of $1.81~\mathrm{THz}$ at $300~\mathrm{K}$ and $1.4~\mathrm{THz}$ at $20~\mathrm{K}$, respectively~\cite{perry_PhysRevB.39.8666}. At intermediate temperatures, the TA mode frequency is obtained by linear interpolation between these two experimental values.
\clearpage

\section{Temperature dependent three-dimensional differential pair distribution function}
\label{sec:3dpdf}

\begin{figure}[htb]
\begin{center}
  {\includegraphics[width=18cm]{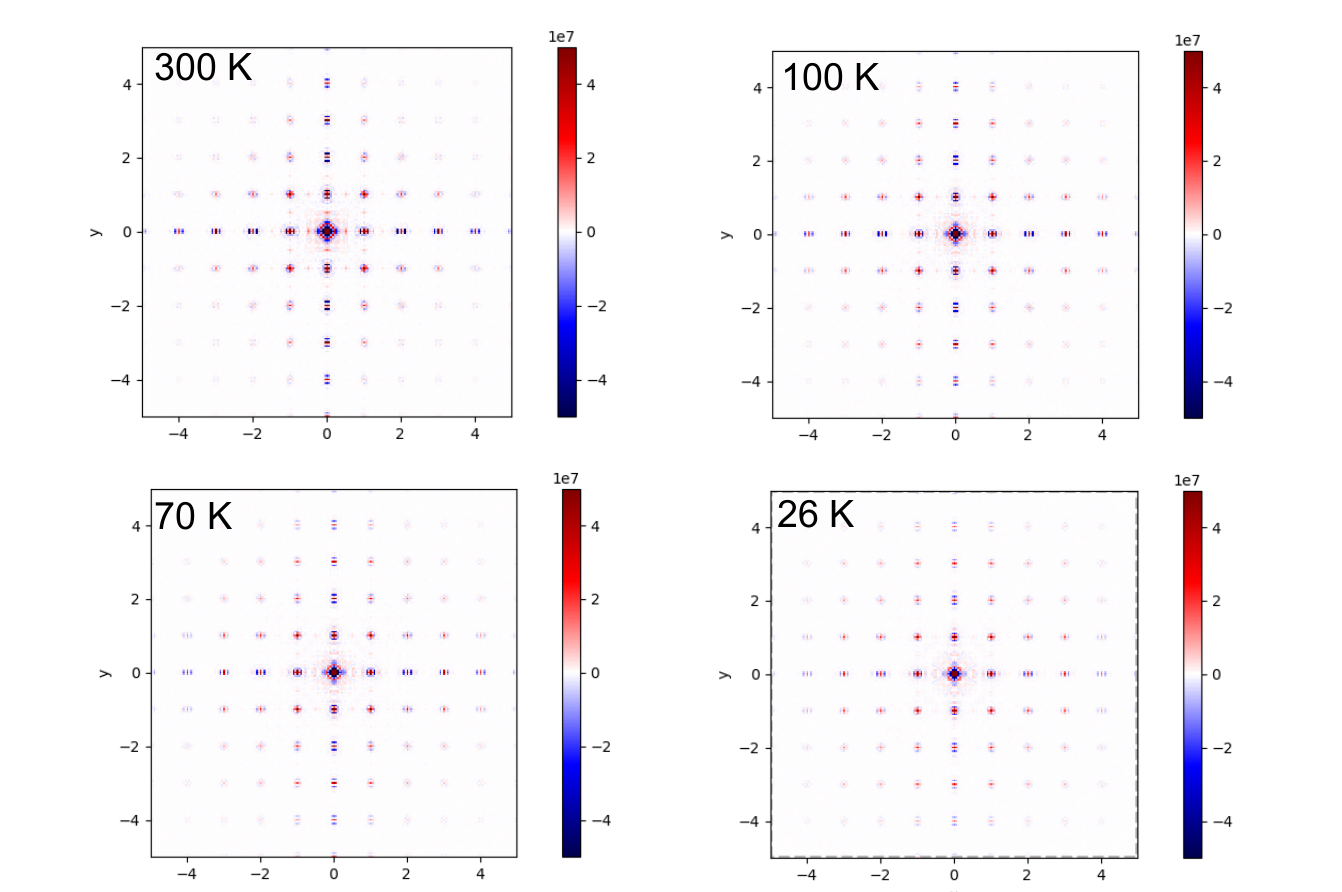}}
  \caption{Temperature evolution of the three-dimensional differential pair distribution function in KTaO$_3$. 3D-$\Delta$PDF maps reconstructed from diffuse X-ray scattering data at $300~\mathrm{K}$, $100~\mathrm{K}$, $70~\mathrm{K}$ and $26~\mathrm{K}$. The overall real-space correlation pattern remains qualitatively unchanged upon cooling, while its intensity progressively decreases.
  }
  \label{sup_detaPDF}
\end{center}
\end{figure}

To obtain a real-space representation of the temperature-dependent two-body correlations in KTaO$_3$, we calculated the three-dimensional differential pair distribution function (3D-$\Delta$PDF) from the diffuse X-ray scattering data using the `punch-and-fill' method~\cite{weber2012three}, as shown in Fig.~\ref{sup_detaPDF}. Across the investigated temperature range from $300~\mathrm{K}$ to $26~\mathrm{K}$, the overall 3D-$\Delta$PDF pattern remains qualitatively unchanged, while its intensity progressively decreases upon cooling. The maps exhibit positive features centered near average interatomic vectors together with negative features at neighboring positions. This characteristic sign pattern is consistent with correlated, predominantly in-phase atomic displacements expected for acoustic-phonon-like fluctuations. Combined with the results of reciprocal space diffuse X-ray scattering in the main text, these real-space correlations support a picture in which the diffuse scattering is dominated by acoustic phonons throughout the investigated temperature range, with an additional contribution from the soft polar mode superimposed at elevated temperatures.

\clearpage

\section{Theoretical calculation of the electronic SHG}
\label{sec:calculations}
The observed SHG response is governed by electronic second-order nonlinear susceptibility. In a centrosymmetric crystal, however, the second-order nonlinear susceptibility vanishes, $\chi^{(2)}_{ijk}=0$. Thus, direct virtual transitions in KTaO$_3$ between the valence and conduction bands alone cannot account for the observed SHG signal within the electric-dipole channel. Further more, any intermediate extended Bloch orbitals based purely on the ideal cubic lattice would still obey the symmetry constraint and cannot generate a nonzero$\chi^{(2)}_{ijk}$. These considerations rule out conventional interband pathways involving pure crystal orbitals in KTaO$_3$. Instead, a defect state that locally breaks inversion symmetry provides a natural intermediate channel with finite $\chi^{(2)}_{ijk}$ for the virtual transition underlying the second-order nonlinear response. Therefore, the dominant microscopic process can be viewed as a defect-assisted virtual transition
\begin{equation}
    |v,\mathbf{k}\rangle
    \rightarrow
    |d_0\rangle
    \rightarrow
    |c,\mathbf{k}\rangle
    \rightarrow
    |v,\mathbf{k}\rangle,
\end{equation}
where $|d_0\rangle$ denotes a localized in-gap defect state.

Importantly, the defect-assisted process discussed here is a virtual intermediate-state process~\cite{sakurai2020modern} named parametric process rather than a sequence of real optical excitations. 
The localized defect level therefore acts as an intermediate nonlinear channel that mediates the effective two-photon coupling between the valence and conduction bands. 
As a consequence, energy conservation is not required at each individual transition step.

\subsection{$~~~$Light-matter interaction and defect-assisted optical transitions}
\label{subsec:light matter interaction}
Within the dipole approximation, the light-matter interaction Hamiltonian is
\begin{equation}
    H_{\rm int}
    =
    -\hat{\mathbf d}\cdot \mathbf E,
\end{equation}
with $\hat{\mathbf d}$ the dipole operator and 
 $\hat{\mathbf e}$ the polarization direction.  The projected dipole matrix elements are defined as~\cite{Haug1994,kittel1963quantum}
\begin{equation}
    d_{mn}(\mathbf{k})
    =
    \langle m,\mathbf{k}|
    \hat{\mathbf d}\cdot \hat{\mathbf e}
    |n,\mathbf{k}\rangle .
\end{equation}

Restricting to the minimal subspace
\begin{equation}
\{
|v,\mathbf{k}\rangle,
|d_0\rangle,
|c,\mathbf{k}\rangle
\},
\end{equation}
the effective Hamiltonian becomes
\begin{equation}
    H
    =
    \begin{pmatrix}
        \varepsilon_v(\mathbf{k})
        &
        V_{vd}(\mathbf{k})
        &
        0
        \\
        V_{dv}(\mathbf{k})
        &
        \varepsilon_d
        &
        V_{dc}(\mathbf{k})
        \\
        0
        &
        V_{cd}(\mathbf{k})
        &
        \varepsilon_c(\mathbf{k})
    \end{pmatrix},
    \label{eq:Hnondiag_static}
\end{equation}
where
\begin{equation}
    V_{mn}(\mathbf{k})
    =
    -d_{mn}(\mathbf{k})E.
\end{equation}
Here $\varepsilon_v(\mathbf{k})$ and $\varepsilon_c(\mathbf{k})$ denote the valence- and conduction-band dispersions, respectively, while $\varepsilon_d$ is the localized defect-state energy. 
Direct dipole transitions between the valence and conduction bands are neglected in this minimal description, such that the optical process is dominated by the virtual defect-assisted channel.

\subsection{$~~~$Schrieffer-Wolff derivation of the effective nonlinear optical coupling}
\label{subsec:Schrieffer Wolff}
We now derive the effective defect-assisted nonlinear coupling by perturbatively eliminating the localized defect state using a Schrieffer-Wolff (SW) transformation~\cite{PhysRev.149.491}. 
The Hamiltonian in Eq.~(\ref{eq:Hnondiag_static}) is decomposed into a diagonal part and an off-diagonal perturbation,
\begin{equation}
    H=H_0+V,
\end{equation}
with
\begin{equation}
    H_0=
    \begin{pmatrix}
        \varepsilon_v(\mathbf{k}) & 0 & 0 \\
        0 & \varepsilon_d & 0 \\
        0 & 0 & \varepsilon_c(\mathbf{k})
    \end{pmatrix},
\end{equation}
and
\begin{equation}
    V=
    \begin{pmatrix}
        0 & V_{vd}(\mathbf{k}) & 0 \\
        V_{dv}(\mathbf{k}) & 0 & V_{dc}(\mathbf{k}) \\
        0 & V_{cd}(\mathbf{k}) & 0
    \end{pmatrix}.
\end{equation}

The purpose of the SW transformation is to integrate out the intermediate defect level and derive an effective Hamiltonian acting only within the low-energy band subspace
\begin{equation}
\{
|v,\mathbf{k}\rangle,
|c,\mathbf{k}\rangle
\}.
\end{equation}

We introduce a unitary transformation
\begin{equation}
    H_{\rm eff}
    =
    e^{S}He^{-S},
\end{equation}
where the generator $S$ is anti-Hermitian,
\begin{equation}
    S^\dagger=-S.
\end{equation}

Expanding the transformed Hamiltonian perturbatively gives
\begin{align}
    H_{\rm eff}
    &=
    H
    +
    [S,H]
    +
    \frac12[S,[S,H]]
    +
    O(V^3)
   =
    H_0
    +
    V
    +
    [S,H_0]
    +
    [S,V]
    +
    \frac12[S,[S,H_0]]
    +
    O(V^3).
\end{align}

To eliminate the first-order coupling between the defect state and the band states, the generator $S$ is chosen such that
\begin{equation}
    [S,H_0]=-V.
\end{equation}
Under this condition, the effective Hamiltonian reduces to
\begin{equation}
    H_{\rm eff}
    =
    H_0
    +
    \frac12[S,V]
    +
    O(V^3).
\end{equation}

The matrix elements of $S$ satisfy
\begin{equation}
    (\varepsilon_n-\varepsilon_m)S_{mn}
    =
    -V_{mn},
\end{equation}
which gives
\begin{equation}
    S_{mn}
    =
    \frac{V_{mn}}
    {\varepsilon_m-\varepsilon_n},
    \qquad
    m\neq n.
\end{equation}

For the present three-level system, the nonzero components are
\begin{align}
    S_{vd}
    =
    \frac{V_{vd}}
    {\varepsilon_v-\varepsilon_d},
    \qquad
    S_{dv}
    =
    \frac{V_{dv}}
    {\varepsilon_d-\varepsilon_v},
    \qquad
    S_{dc}
    =
    \frac{V_{dc}}
    {\varepsilon_d-\varepsilon_c},
    \qquad
    S_{cd}
    =
    \frac{V_{cd}}
    {\varepsilon_c-\varepsilon_d}.
\end{align}

The effective coupling between the valence and conduction bands arises from the second-order commutator term. 
Taking the $(v,c)$ matrix element gives
\begin{align}
    \left(H_{\rm eff}\right)_{vc}^{(2)}
    &=
    \frac12[S,V]_{vc}
   =
    \frac12
    \sum_m
    \left(
    S_{vm}V_{mc}
    -
    V_{vm}S_{mc}
    \right).
\end{align}

Since the defect level is the only intermediate state retained in the minimal model, only $m=d$ contributes,
\begin{equation}
    \left(H_{\rm eff}\right)_{vc}^{(2)}
    =
    \frac12
    \left(
    S_{vd}V_{dc}
    -
    V_{vd}S_{dc}
    \right).
\end{equation}

Substituting the explicit expressions for $S_{vd}$ and $S_{dc}$ yields
\begin{align}
    \left(H_{\rm eff}\right)_{vc}^{(2)}
    &=
    \frac12
    V_{vd}V_{dc}
    \left[
    \frac{1}{\varepsilon_v-\varepsilon_d}
    -
    \frac{1}{\varepsilon_d-\varepsilon_c}
    \right]
    =
    \frac12
    V_{vd}V_{dc}
    \left[
    \frac{1}{\varepsilon_v-\varepsilon_d}
    +
    \frac{1}{\varepsilon_c-\varepsilon_d}
    \right].
    \label{eq:Heff_SW_static}
\end{align}

Using
\begin{equation}
    V_{mn}
    =
    -d_{mn}E,
\end{equation}
the effective second-order coupling becomes
\begin{equation}
    \left(H_{\rm eff}\right)_{vc}^{(2)}
  =
    \frac{1}{2}d_{vd}d_{dc}E^2 \left[
    \frac{1}{\varepsilon_v-\varepsilon_d}
    +
    \frac{1}{\varepsilon_c-\varepsilon_d}
    \right].
\end{equation}

Therefore, integrating out the localized defect state generates an effective nonlinear optical coupling between the valence and conduction bands that is quadratic in the electric field. 
The coupling strength is controlled by the product of dipole matrix elements and the detuning from the intermediate defect level. 
Because inversion symmetry is locally broken around the defect environment, the resulting defect-assisted electronic process naturally produces a finite second-order susceptibility and hence a nonzero SHG response even when the bulk crystal remains globally centrosymmetric.\\

\subsection{$~~~$Parity constraint on dipole transition matrix elements}
\label{subsec:parity constraint}
The existence of a finite second-order nonlinear optical response is closely related to the parity structure of the dipole transition matrix elements. 
In the electric-dipole approximation, the optical coupling operator is
\begin{equation}
    H_{\rm int}
    =
    -\hat{\mathbf d}\cdot \mathbf E,
\end{equation}
where the dipole operator is odd under spatial inversion,
\begin{equation}
    \mathcal P \hat{\mathbf d}\mathcal P^{-1}
    =
    -\hat{\mathbf d}.
\end{equation}
Consequently, the dipole matrix element between two eigenstates depends strongly on their inversion parity.

Suppose that the electronic eigenstates possess well-defined parity quantum numbers,
\begin{equation}
    \mathcal P |n\rangle
    =
    p_n |n\rangle,
    \qquad
    p_n=\pm1.
\end{equation}
The dipole matrix element is
\begin{equation}
    d_{mn}
    =
    \langle m|
    \hat{\mathbf d}
    |n\rangle .
\end{equation}
Applying inversion symmetry gives
\begin{align}
    d_{mn}
    &=
    \langle m|
    \mathcal P^{-1}
    \mathcal P
    \hat{\mathbf d}
    \mathcal P^{-1}
    \mathcal P
    |n\rangle=
    -p_m p_n
    \langle m|
    \hat{\mathbf d}
    |n\rangle
    =
    -p_m p_n d_{mn}.
\end{align}
Therefore,
\begin{equation}
    (1+p_m p_n)d_{mn}=0.
\end{equation}
A nonzero dipole transition matrix element requires
\begin{equation}
    p_m p_n=-1,
\end{equation}
meaning that electric-dipole transitions are only allowed between states of opposite parity.

This parity selection rule has important consequences for second-harmonic generation. 
For a purely centrosymmetric bulk crystal, the valence and conduction Bloch states possess definite inversion parity, and the nonlinear optical susceptibility must satisfy
 $\chi^{(2)}_{ijk}=0$. As a result, a direct two-photon process involving only bulk valence and conduction states cannot generate a finite electric-dipole SHG response.

Localized defect states modify this situation in two important ways. 
First, the defect potential locally breaks inversion symmetry, so the corresponding electronic wavefunctions no longer possess definite parity eigenvalues. 
Second, the defect state generally contains admixtures of both even- and odd-parity orbital components,
\begin{equation}
    |d_0\rangle
    =
    \alpha |+\rangle
    +
    \beta |-\rangle ,
\end{equation}
where $|+\rangle$ and $|-\rangle$ denote even- and odd-parity components, respectively. 
Because of this parity mixing, both dipole matrix elements
\begin{equation}
    d_{vd}
    =
    \langle v|
    \hat{\mathbf d}\cdot{\bf e}
    |d_0\rangle ,\qquad
    d_{dc}
    =
    \langle d_0|
    \hat{\mathbf d}\cdot{\bf e}
    |c\rangle , 
\end{equation}
can simultaneously become finite even when the direct bulk transition matrix element between the valence and conduction bands is symmetry forbidden.  In particular, if $v$ and $c$ have opposite parity, the product generally involves both parity components and may scale as $d_{vd}d_{dc}\propto \alpha\beta$. Thus, the parity mixing in the defect state allows both defect-assisted dipole matrix elements to become simultaneously nonzero. In intrinsically non-centrosymmetric materials, where inversion symmetry is already broken at the bulk crystal level, such parity mixing is expected to be even stronger and can therefore generate substantially enhanced electric-dipole second-order nonlinear optical responses.

The defect-assisted virtual process
\begin{equation}
    |v,\mathbf{k}\rangle
    \rightarrow
    |d_0\rangle
    \rightarrow
    |c,\mathbf{k}\rangle
    \rightarrow
    |v,\mathbf{k}\rangle
\end{equation}
therefore bypasses the parity restriction that suppresses the bulk electric-dipole SHG response. 
After integrating out the intermediate defect level, the resulting effective nonlinear coupling is proportional to
\begin{equation}
    H_{\rm eff}^{(2)}
    \propto
    d_{vd}d_{dc}E^2,
\end{equation}
which produces a finite second-order susceptibility. 
Thus, the localized defect state acts both as a virtual intermediate channel and as a local inversion-symmetry-breaking source that enables electric-dipole SHG in an otherwise centrosymmetric material.

\subsection{$~~~$Soft-phonon modulation of the coherent nonlinear vertex}
\label{subsec:nonlinear vertex}
The defect-assisted SHG process is strongly influenced by the fluctuating polar environment surrounding the localized defect state. 
In particular, the soft transverse optical phonons in KTaO$_3$ dynamically modulate the local inversion-symmetry-breaking field and therefore modify the coherent nonlinear optical vertex associated with the defect-assisted transition channel.

We define the effective defect-assisted nonlinear transition vertex as
\begin{equation}
    M_{cvd}(\mathbf{k})
    =
    \langle c,\mathbf{k}|
    \hat{\mathbf d}\cdot\hat{\mathbf e}
    |d_0\rangle
    \langle d_0|
    \hat{\mathbf d}\cdot\hat{\mathbf e}
    |v,\mathbf{k}\rangle .
\end{equation}
This quantity enters directly into the effective second-order optical coupling derived in the previous section.

Because the localized defect state is embedded in a fluctuating polar lattice environment, the corresponding dipole matrix elements become dependent on the local ionic configuration. 
Introducing the phonon normal coordinates $\{Q_\nu\}$, the nonlinear vertex becomes configuration dependent,
\begin{equation}
    M_{cvd}(\mathbf{k};\{Q_\nu\})
    =
    \langle c,\mathbf{k}|
    \hat{\mathbf d}\cdot\hat{\mathbf e}
    |d[\{Q_\nu\}]\rangle
    \langle d[\{Q_\nu\}]|
    \hat{\mathbf d}\cdot\hat{\mathbf e}
    |v,\mathbf{k}\rangle .
    \label{eq:SMMQ}
\end{equation}

The experimentally observed SHG signal probes the coherent thermal average of this nonlinear amplitude,
\begin{equation}
    \overline{M}_{cvd}(\mathbf{k},T)
    =
    \left\langle
    M_{cvd}(\mathbf{k};\{Q_\nu\})
    \right\rangle_T .
    \label{eq:SMMTavg}
\end{equation}

For weak lattice fluctuations, the dominant effect of the soft polar environment is to induce fluctuating phases in the defect-assisted nonlinear polarization channel through local inversion- and translational-symmetry-breaking distortions around the defect configuration. We therefore write
\begin{equation}
    M_{cvd}(\mathbf{k};\{Q_\nu\})
    =
    M_{cvd}(\mathbf{k})
    e^{-i\phi(\{Q_\nu\})},
    \label{eq:SMphaseform}
\end{equation}
where the fluctuating phase field is expanded as $
    \phi(\{Q_\nu\})
    =
    \sum_\nu g_\nu Q_\nu$. Here $g_\nu$ denotes the effective coupling between phonon mode $\nu$ and the coherent nonlinear optical channel. For harmonic phonons, the normal coordinates obey Gaussian thermal fluctuations. 
Using a second-order cumulant expansion~\cite{peskin2018introduction},
\begin{equation}
    \left\langle
    e^{-i\phi}
    \right\rangle_T
    =
    \exp
    \left[
    -\frac12
    \left\langle
    \phi^2
    \right\rangle_T
    \right],
\end{equation}
the coherent nonlinear vertex becomes
\begin{equation}
    \overline{M}_{cvd}(\mathbf{k},T)
    =
    M_{cvd}(\mathbf{k})
    e^{-W(T)},
    \label{eq:SMMrenorm}
\end{equation}
with the Debye-Waller-type suppression factor
\begin{equation}
    W(T)
    =
    \frac12
    \sum_\nu
    |g_\nu|^2
    \left\langle
    Q_\nu^2
    \right\rangle_T .
    \label{eq:SMWgeneral}
\end{equation}

For a harmonic phonon mode of frequency $\Omega_\nu$~\cite{abrikosov2012methods,mahan2013many,kittel1963quantum},
\begin{equation}
    \left\langle
    Q_\nu^2
    \right\rangle_T
    =
    \frac{\hbar}{2\Omega_\nu}
    \left[
    2n_B(\Omega_\nu,T)+1
    \right],
\end{equation}
where $
    n_B(\Omega_\nu,T)$ is the Bose distribution function.

Because the transverse optical soft mode in KTaO$_3$ possesses an anomalously small energy scale, its displacement fluctuations become strongly enhanced. 
As a consequence, the soft mode gives the dominant contribution to the phase fluctuation factor $W(T)$ and can strongly suppress the coherent defect-assisted nonlinear optical vertex at elevated temperature.

\subsection{$~~~$Semiconductor Bloch-equation description of the SHG response}
\label{subsec:Bloch equation}

We now formulate the defect-assisted SHG response using a semiconductor
Bloch-equation description~\cite{Haug1994,Wu2010SpinDynamics,kittel1963quantum}. The central microscopic variable is the interband
coherence $\rho_{cv}(\mathbf{k},t)$, which describes the quantum coherence
between the valence-band state $|v,\mathbf{k}\rangle$ and the conduction-band
state $|c,\mathbf{k}\rangle$. The diagonal density-matrix elements
$\rho_{vv}(\mathbf{k},t)$ and $\rho_{cc}(\mathbf{k},t)$ describe the band
occupations, while the off-diagonal component $\rho_{cv}(\mathbf{k},t)$
describes the optical polarization coherence.

The macroscopic polarization radiating at the second-harmonic frequency is
obtained by summing the microscopic interband coherences over momentum~\cite{Haug1994,Wu2010SpinDynamics,kittel1963quantum},
\begin{equation}
    P^{(2)}(2\omega;T)
    =
    \sum_{\mathbf{k}}
    d_{vc}(\mathbf{k})
    \rho_{cv}^{(2)}(\mathbf{k},2\omega;T),
    \label{eq:P_from_pcv}
\end{equation}
where $
    d_{vc}(\mathbf{k})
    =
    \langle v,\mathbf{k}|
    \hat{\mathbf d}\cdot \hat{\mathbf e}
    |c,\mathbf{k}\rangle$ is the dipole matrix element associated with the emitted second-harmonic
polarization.

In the relaxation-time approximation, the equation of motion for the interband
coherence is written as~\cite{Haug1994,Wu2010SpinDynamics,kittel1963quantum}
\begin{equation}
    i\hbar
    \frac{d}{dt}
    \rho_{cv}(\mathbf{k},t)
    =
    \left[
    \varepsilon_c(\mathbf{k})
    -
    \varepsilon_v(\mathbf{k})
    -
    i\hbar \gamma_{\mathbf{k}}(T)
    \right]
    \rho_{cv}(\mathbf{k},t)
    +
    \mathcal{D}_{cv}(\mathbf{k})
    \mathcal{F}_{cv}^{(2)}(\mathbf{k},T)
    e^{-2i\omega t}.
    \label{eq:SBE_pcv}
\end{equation}
Here $
    \mathcal{D}_{cv}(\mathbf{k})
    =
    \rho_{vv}(\mathbf{k})-\rho_{cc}(\mathbf{k})$ 
is the equilibrium occupation difference. For an insulating system under weak
optical excitation,
\begin{equation}
    \rho_{vv}(\mathbf{k})=f_v(\mathbf{k})\simeq 1,
    \qquad
    \rho_{cc}(\mathbf{k})=f_c(\mathbf{k})\simeq 0,
    \qquad
    \mathcal{D}_{cv}(\mathbf{k})\simeq 1 .
\end{equation}

The effective second-order driving vertex derived in the previous sections is
\begin{equation}
    \mathcal{F}_{cv}^{(2)}(\mathbf{k},T)
    =
    M_{cvd}(\mathbf{k})e^{-W(T)} 
    \Lambda(\mathbf{k})
    E_\omega^2 ,
    \label{eq:Fcv_def}
\end{equation}
where $
    M_{cvd}(\mathbf{k})
    =
    d_{cd}(\mathbf{k})d_{dv}(\mathbf{k})$  
is the product of the two defect-assisted dipole matrix elements; the factor $
    \Lambda(\mathbf{k})
    =
    \frac{1}{2}
[
    \frac{1}{\varepsilon_v(\mathbf{k})-\varepsilon_d}
    +
    \frac{1}{\varepsilon_c(\mathbf{k})-\varepsilon_d}
]$ is the virtual-state energy denominator obtained from the Schrieffer-Wolff
transformation; the factor $e^{-W(T)}$ is the Debye-Waller-like coherence factor derived in the
previous subsection. This factor describes the loss of phase coherence of the
defect-assisted nonlinear polarization channel due to fluctuating local polar
distortions. The quantity $\gamma_{\mathbf{k}}(T)$ in Eq.~(\ref{eq:SBE_pcv}) is the
inhomogeneous dephasing rate of the interband coherence~\cite{Haug1994,Wu2010SpinDynamics,kittel1963quantum}. It enters the electronic
resonance denominator as a linewidth and should be distinguished from the
coherent vertex suppression factor $e^{-W(T)}$.

We seek the second-harmonic component of the interband coherence in the form
\begin{equation}
    \rho_{cv}^{(2)}(\mathbf{k},t)
    =
    \rho_{cv}^{(2)}(\mathbf{k},2\omega)
    e^{-2i\omega t}.
\end{equation}
Substituting this ansatz into Eq.~(\ref{eq:SBE_pcv}) gives
\begin{equation}
    \rho_{cv}^{(2)}(\mathbf{k},2\omega;T)
    =
    \frac{
    \mathcal{D}_{cv}(\mathbf{k})
    \overline{M}_{cvd}(\mathbf{k},T)
    \Lambda(\mathbf{k})
    E_\omega^2
    }{
    \varepsilon_c(\mathbf{k})
    -
    \varepsilon_v(\mathbf{k})
    -
    2\hbar\omega
    -
    i\hbar\gamma_{\mathbf{k}}(T)
    } .
    \label{eq:pcv_solution}
\end{equation}
The denominator describes the detuning between the two-photon energy and the
effective interband transition energy, with Lorentzian broadening controlled by
the dephasing rate.

Inserting Eq.~(\ref{eq:pcv_solution}) into Eq.~(\ref{eq:P_from_pcv}), one
obtains
\begin{equation}
    P^{(2)}(2\omega;T)
    =
    E_\omega^2
    \sum_{\mathbf{k}}
    \frac{
    d_{vc}(\mathbf{k})
    \mathcal{D}_{cv}(\mathbf{k})
    M_{cvd}(\mathbf{k})
    e^{-W(T)}
    \Lambda(\mathbf{k})
    }{
    \varepsilon_c(\mathbf{k})
    -
    \varepsilon_v(\mathbf{k})
    -
    2\hbar\omega
    -
    i\hbar\gamma_{\mathbf{k}}(T)
    } .
    \label{eq:P2omega_final}
\end{equation}
The experimentally detected second-harmonic intensity is proportional to the
squared modulus of the coherent macroscopic polarization,
\begin{equation}
    I_{2\omega}(T)
    \propto
    \left|
    P^{(2)}(2\omega;T)
    \right|^2 .
\end{equation}
Therefore,
\begin{equation}
    I_{2\omega}(T)
    \propto
    |E_\omega|^4
    \left|
    \sum_{\mathbf{k}}
    \frac{
    d_{vc}(\mathbf{k})
    \mathcal{D}_{cv}(\mathbf{k})
    M_{cvd}(\mathbf{k})
    e^{-W(T)}
    \Lambda(\mathbf{k})
    }{
    \varepsilon_c(\mathbf{k})
    -
    \varepsilon_v(\mathbf{k})
    -
    2\hbar\omega
    -
    i\hbar\gamma_{\mathbf{k}}(T)
    }
    \right|^2 .
    \label{eq:I2omega_final}
\end{equation}
This expression separates two distinct temperature-dependent effects. First,
soft polar fluctuations suppress the coherent nonlinear vertex through the
factor $e^{-W(T)}$, which reduces the phase coherence of the defect-assisted SHG
channel. Second, phonon and disorder scattering broaden the electronic resonance
through $\gamma_{\mathbf{k}}(T)$, thereby reducing the resonant enhancement of
the interband coherence.\\

The dephasing rate is modeled phenomenologically as
\begin{equation}
    \gamma_{\mathbf{k}}(T)
    =
    \gamma_{0,\mathbf{k}}
    +
    \gamma_{{\rm soft},\mathbf{k}}(T)
    +
    \gamma_{{\rm ac},\mathbf{k}}(T),
    \label{eq:SMgamma}
\end{equation}
where $\gamma_{0,\mathbf{k}}$ is the residual defect and disorder broadening,
$\gamma_{{\rm soft},\mathbf{k}}(T)$ originates from thermally populated soft
polar modes, and $\gamma_{{\rm ac},\mathbf{k}}(T)$ is the acoustic-phonon
contribution. The acoustic contribution  accounts for an approximate $T^3$
linewidth dependence, while the dominant suppression of the coherent SHG
intensity is controlled by the soft-mode Debye-Waller factor $e^{-W(T)}$.

Since the SHG field is a coherent sum over microscopic polarization amplitudes,
thermal phase fluctuations suppress the amplitude before taking the squared
modulus. Thus, when the soft-mode coherence factor is weakly momentum dependent,
Eq.~(\ref{eq:I2omega_final}) implies approximately
\begin{equation}
    I_{2\omega}(T)
    \sim
    e^{-2W(T)}
    I_{2\omega}^{(0)}(T),
\end{equation}
where $I_{2\omega}^{(0)}(T)$ contains the remaining electronic resonance and
linewidth effects. The dominant temperature dependence of the SHG intensity originates
from the soft-phonon-induced coherence factor $e^{-2W(T)}$. Physically, thermal
fluctuations of the soft polar mode progressively randomize the phase of the
defect-assisted nonlinear polarization channel, thereby suppressing the
coherent second-harmonic response. In contrast, the linewidth contribution
$\gamma_{\mathbf{k}}(T)$ mainly produces additional spectral broadening and
only provides a subleading correction to the overall temperature dependence of
the SHG intensity. 

The theoretical fitting of the experimentally observed SHG intensity  in the main text is primarily based on the temperature dependence of the soft-mode Debye-Waller factor $e^{-2W(T)}$ derived above. 
Assuming an isotropic soft-mode dispersion,
\begin{equation}
\Omega_P(q)
=
\sqrt{
\Delta^2(T)
+
v_s^2 q^2
},
\end{equation}
where $\Delta(T)=\hbar \omega_{q=0}(T)$ is the temperature-dependent zone-center soft-mode gap, taken from the values reported in Ref.~\citenum{74d5-4hsw}.
In that work, $\omega_q(T)$ was obtained from the self-consistent soft-phonon renormalization model, and the resulting zone-center frequencies were shown to agree well with the experimentally extracted soft-mode frequencies.
The mode velocity $v_s$ is adopted from the same parameter set, where it was determined by comparison with low-temperature inelastic neutron and Raman scattering data~\cite{rowley2014ferroelectric}.  Within this continuum approximation, the soft-mode contribution to the Debye-Waller fluctuation factor~\cite{74d5-4hsw,gpbp-qhp9}
\begin{equation}
    W_{\rm soft}(T)
    =
    \frac{\eta}{2\pi^2}
    \int_{0}^{q_D}
    q^2dq\,
    \frac{
    2n_B(\Omega_P(q),T)+1
    }{
    2\Omega_P(q)
    },
    \label{eq:SMWsoft}
\end{equation}
where $q_D$ is the effective isotropic momentum cutoff for the soft-mode fluctuation spectrum.
Its precise value mainly rescales the fluctuation amplitude and is therefore absorbed into the fitted coupling $\eta$, which also collects the effective coupling constants, and normalization factors required to make $W_{\rm soft}(T)$ dimensionless.
All other quantities entering Eq.~\eqref{eq:SMWsoft}, including the temperature-dependent zone-center soft-mode gap and the mode velocity, are fixed from the soft-mode parameterization of Ref.~\cite{74d5-4hsw}. 
Therefore, for the normalized SHG intensity $I_{\rm SHG}(T)/I_{\rm SHG}(T_0)$, $\eta$ is the only adjustable parameter controlling the Debye-Waller suppression, with the best-fit value $\eta=18~{\rm eV\,\AA^3}$ in the convention where $q$ is measured in ${\rm \AA^{-1}}$ and $\Omega_P$ in eV.. As shown in the main text, after fitting the single parameter $\eta$, the calculated temperature dependence of the normalized SHG intensity exhibits remarkably good agreement with the experimental data over the entire measured temperature range. The linewidth parameters were treated separately: the residual broadening was fixed to $\gamma_0=87.5~{\rm meV}$ according to the low-temperature FWHM scale, and the weak temperature-dependent linewidth correction was described using $c_0=2.0\times10^{-8}$ in the same unit convention. This consistency strongly supports the interpretation that the observed SHG suppression is predominantly governed by thermally enhanced soft-mode fluctuations through the Debye-Waller mechanism.

Consequently, the resulting suppression factor $\exp{[-W_{\rm soft}(T)]}$ acts as a coherence factor for the defect-assisted nonlinear optical amplitude.
Physically, increasing soft-phonon fluctuations progressively randomize the local phase of the defect-induced nonlinear polarization, thereby reducing the coherent SHG response even though the local inversion symmetry breaking itself remains finite. 
This mechanism naturally explains the strong temperature dependence of the experimentally observed SHG intensity in the soft-phonon regime of KTaO$_3$.

\subsection{$~~~$First-principles calculations}
\label{subsec:first principles}

The first-principles calculations were performed using the Vienna \textit{ab-initio} simulation package (VASP)~\cite{Kresse1996-1,Kresse1996-2}.
The projector-augmented-wave (PAW) method was employed together with a plane-wave cutoff energy of 600~eV. 
Spin-polarized structural relaxations and electronic-structure calculations for oxygen-defective KTaO$_3$ were primarily carried out using the screened hybrid functional of Heyd, Scuseria, and Ernzerhof (HSE06)~\cite{HSE06-1,HSE06-2,HSE06-3}, which provides an improved description of the band gap and localized defect states compared with semilocal density functionals.

To model an isolated oxygen vacancy, we constructed a $2\times2\times4$ KTaO$_3$ supercell containing a single oxygen vacancy. 
The Brillouin zone was sampled using a $2\times2\times1$ $k$-point mesh.
All atomic positions were fully relaxed until the residual force on each atom was smaller than 5~meV/\AA, while the electronic self-consistent loop was converged to an energy tolerance of $10~\mu$eV.

The singly charged oxygen vacancy, $V_{\rm O}^{+}$, was modeled by removing one electron from the neutral defective supercell together with a homogeneous compensating background charge in order to avoid the divergence of the electrostatic energy under periodic boundary conditions. 
This charge state was chosen based on previous hybrid-functional studies of intrinsic defects in KTaO$_3$~\cite{modak2021energetic}. 
According to that work, the $V_{\rm O}^{+}$ configuration retains an isolated in-gap defect state with the Fermi level positioned between the defect level and the conduction band, and remains thermodynamically stable over a finite range of Fermi-level positions near the upper part of the band gap. 
By contrast, the neutral vacancy $V_{\rm O}^{0}$ leads to electron occupation in the conduction band, while the doubly charged vacancy $V_{\rm O}^{2+}$ does not generate a well-isolated in-gap defect level~\cite{modak2021energetic}.

We therefore use $V_{\rm O}^{+}$ as a representative optically active oxygen vacancy configuration relevant to the defect-assisted SHG mechanism discussed in this work. 
In our HSE06 calculations, the $2\times2\times4$ supercell containing $V_{\rm O}^{+}$ exhibits a localized and weakly dispersive in-gap defect band. 
The elongated supercell geometry substantially reduces the artificial interaction between periodic images of the vacancy compared with smaller supercells, thereby yielding a more localized defect state appropriate for modeling virtual defect-assisted optical transitions.

At elevated oxygen-vacancy concentrations, vacancy-vacancy interactions and clustering effects may become relevant. 
Motivated by previous theoretical work showing that the linear oxygen divacancy is one of the energetically favorable divacancy configurations in KTaO$_3$~\cite{Ojha2021divac}, we also explored this defect geometry. 
Spin-polarized atomic and electronic structures for a $2\times2\times4$ KTaO$_3$ supercell containing a linear oxygen divacancy were calculated within the generalized-gradient approximation using the PBEsol exchange-correlation functional~\cite{Perdew2008}. 
The Brillouin zone was sampled using a $4\times4\times2$ $\Gamma$-centered $k$-point grid. 
The atomic positions were fully relaxed until the residual force on each atom was smaller than 5~meV/\AA, and the electronic self-consistent loop was converged to an energy tolerance of $0.1~\mu$eV.

\subsection{$~~~$Construction of the momentum-resolved defect-assisted nonlinear vertex}
\label{subsec:construction of nonlinear vertex}

To connect the first-principles electronic structure with the effective
nonlinear optical theory developed above, we constructed the
momentum-resolved defect-assisted nonlinear transition vertex
$M_{cvd}(\mathbf{k})$ from the calculated dipole transition matrix elements.

The transition dipole moments were extracted from the HSE06 wave functions
using the \texttt{kit} transition-dipole post-processing workflow. In the
defective supercell, band 237 was identified as the localized in-gap defect
state $d_0$. Since the valence-band maximum and conduction-band minimum each
consist of several closely spaced near-edge bands, the defect-assisted optical
strength was not assigned to a single valence-conduction band pair. Instead,
the nonlinear optical channel was constructed by summing over the dominant
near-edge states.

The valence-band manifold included
\begin{equation}
    \mathcal V
    =
    \{234,235,236\},
\end{equation}
while the conduction-band manifold included
\begin{equation}
    \mathcal C
    =
    \{238,239,240\}.
\end{equation}
The component-resolved transition-dipole data were evaluated for the Cartesian projections $x$, $y$, and $z$ in the crystal coordinate system. In the present defect geometry, the $z$ direction corresponds to the nearest-neighbor Ta--$V_{\rm O}$--Ta axis, whereas $x$ and $y$ denote the transverse in-plane directions.  For the near-edge $\Gamma$--$Z$ region considered here, the two transverse components exhibit very similar behavior and both dominate over the longitudinal $z$ component.  Consequently, in constructing the scalar momentum-resolved optical vertex, we choose the strongest in-plane polarization direction, $\hat{\mathbf e}=\hat{\mathbf y}$. Owing to the approximate in-plane symmetry of the near-edge electronic structure in the $\Gamma$--$Z$ region, using the $x$ component instead would lead to essentially the same physical conclusions.

For each momentum point $\mathbf{k}$, we first
constructed the total transition strengths associated with the two legs of
the virtual defect-assisted optical process,
\begin{equation}
    I_{vd}(\mathbf{k})
    =
    \sum_{v\in\mathcal V}
    \left|
    \mu_{v d_0}(\mathbf{k})
    \right|^2,\qquad
    I_{dc}(\mathbf{k})
    =
    \sum_{c\in\mathcal C}
    \left|
    \mu_{d_0 c}(\mathbf{k})
    \right|^2 .
    \label{eq:SMTDMsum}
\end{equation}
Here $
    \mu_{mn}(\mathbf{k})
    =
    \langle m,\mathbf{k}|
    \hat{\mathbf d}\cdot\hat{\mathbf e}
    |n,\mathbf{k}\rangle$  
denotes the dipole transition matrix element projected along the strongest, in-plane optical polarization direction identified above.
From these quantities we defined the effective dipole amplitudes for the two parts of the virtual process,
\begin{equation}
    \mu_{vd,{\rm eff}}(\mathbf{k})
    =
    \sqrt{
    I_{vd}(\mathbf{k})
    },\qquad
    \mu_{dc,{\rm eff}}(\mathbf{k})
    =
    \sqrt{
    I_{dc}(\mathbf{k})
    }.
\end{equation}
The spin-conserving two-step nonlinear transition vertex for each spin channel
was then constructed as
\begin{equation}
    M_{cvd}(\mathbf{k})
    =
    \mu_{dc,{\rm eff}}(\mathbf{k})
    \mu_{vd,{\rm eff}}(\mathbf{k}),
    \label{eq:SMTDMproduct}
\end{equation}
which corresponds to the numerical realization of the effective virtual
transition amplitude 
\begin{equation}
    \langle c,\mathbf{k}|
     \hat{\mathbf d}\cdot\hat{\mathbf e}
    |d_0\rangle
    \langle d_0|
     \hat{\mathbf d}\cdot\hat{\mathbf e}
    |v,\mathbf{k}\rangle .
\end{equation}
The resulting quantity has units of Debye$^2$ and represents the effective
momentum-resolved strength of the defect-assisted two-photon nonlinear optical
channel. Physically, large values of
$|M_{cvd}(\mathbf{k})|$ identify regions of momentum space where
the localized defect state strongly hybridizes optically with both valence and
conduction states, thereby generating an enhanced effective second-order
nonlinear response.

\bibliography{sample}